\documentclass[12pt]{iopart}
\usepackage{graphicx}
\usepackage{color}

\begin{document}

\title{Estimating nonlinear stability from time series data}
\author{Adrian van Kan$^{1,2}$, Jannes Jegminat$^3$, Jonathan F. Donges$^{4,5}$}

\address{$^1$ D\'epartement de Physique de l'\'Ecole Normale Sup\'erieure, Paris, France\\ $^2$ Physics Department, University of California, Berkeley, USA \\ $^3 $  Spiden AG, Pfäffikon SZ, Switzerland \\ $^4$ Earth System Analysis, Potsdam Institute for Climate Impact Research, Member of the Leibniz Association, 14473 Potsdam, Germany \\ $^5$ Stockholm Resilience Centre, Stockholm University, Stockholm, Sweden}
\ead{avankan@berkeley.edu}
\vspace{10pt}
\begin{indented}
\item[] 30 November 2021
\end{indented}

\begin{abstract}
Basin stability (BS) is a measure of nonlinear stability in multistable dynamical systems. BS has previously been estimated using Monte-Carlo simulations, which requires the explicit knowledge of a dynamical model. We discuss the requirements for estimating BS from time series data in the presence of strong perturbations, and illustrate our approach for two simple models of climate tipping elements: the Amazon rain forest and the thermohaline ocean circulation. We discuss the applicability of our method to observational data as constrained by the relevant time scales of total observation time, typical return time of perturbations and internal convergence time scale of the system of interest and other factors.
\end{abstract}

%
%
%
%
%

\section{Introduction}
Over the last decade, there has been a growing interest in quantifying the stability and resilience of states in complex multistable dynamical systems to different kinds of perturbations \cite{lundstrom2018find,tamberg2020guideline}. In the Earth system and its components, multistability occurs at different levels: in atmospheric jets \cite{herbert2020atmospheric,simonnet2021multistability}, between flows with and without large-scale hurricane-like vortices \cite{favier2019subcritical,van2019rare,de2021discontinuous}, in the cryosphere-climate system \cite{calov2005multistability, robinson2012multistability}, and vegetation \cite{kleidon2000assessing,van2014tipping,wuyts2017amazonian,wuyts2019tropical}, among others \cite{lenton2008tipping}. In addition to the classical linear-stability paradigm for studying the response to infinitesimally small perturbations, basin stability $\mathrm{BS}$ has been proposed as a complementary measure, built so as to naturally incorporate finite-amplitude perturbations \cite{menck2013basin}. Instead of relying on local information close to a state of interest, basin stability approximates the size of the basin of attraction to quantify resilience. Basin stability and similar nonlinear measures of resilience have been applied successfully to a variety of different systems, including to the carbon cycle, climate tipping elements such as the Amazon rain forest and tipping cascades \cite{menck2013basin, van2016constrained,wunderling2020basin}, to uncover precursors of transitions that other, more local stability measures based on the linear paradigm \cite{scheffer2009early} cannot detect. Tipping elements can be triggered by large-amplitude fluctuations due to noise \cite{ditlevsen2010tipping}, which makes these nonlinear stability measures highly relevant in this context. \\


 Previous estimations of basin stability and related nonlinear stability measures relied on simulations of system trajectories, i.e., they required explicit knowledge of system's underlying differential equations. However, in many realistic settings, these equations are only approximately or partially known, or not known at all. Instead, one can typically access only time series of one or several system observables. One approach to estimating basin stability would be to first infer the underlying dynamics from time series, using methods from the field of system identification \cite{ljung2020shift, reynders2012system}, and then use the approximated dynamics to simulate trajectories and estimate the basin stability. We propose an alternative and less complex approach: we estimate the basin stability of attractors in multistable dynamical systems directly from the time series. The method is illustrated on two models of climate tipping elements, first a simple model of the Amazon rain forest presented in \cite{van2014tipping}, and second a model of the Atlantic Meridional overturning circulation (AMOC) introduced by Wood et al. \cite{wood2019observable}, and analysed in detail in \cite{alkhayuon2019basin}.

\section{Method}
\label{sec:method}
\begin{figure}
    \centering
    \includegraphics[width=0.5\textwidth]{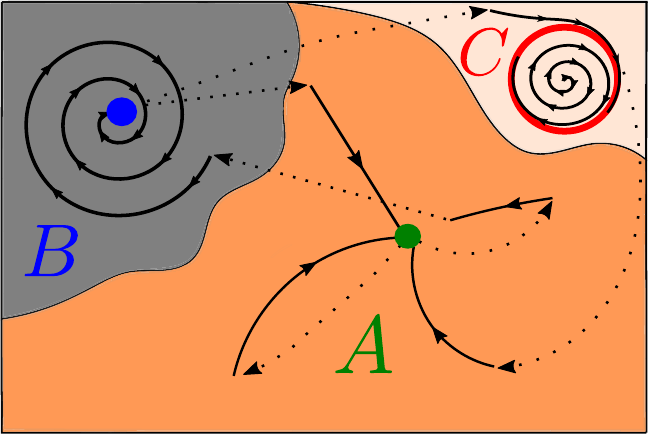}
    \caption{Sketch of the trajectory (solid black lines) in a multistable phase space with three attractors $A,B,C$ subject to perturbations (dashed lines), arrows indicate the direction of the flow. Most perturbations, here drawn from a uniform distribution over the visible state space, take the system to states within the basin of attraction of $A$ which has the largest volume. Trajectories may be disrupted by a perturbation before reaching an attractor (see trajectory to the right of $A$). From the converged trajectories in the example one can estimate $\mathrm{BS}(A)\approx3/5$, $\mathrm{BS}(B)=\mathrm{BS}(C)\approx1/5$.  }
    \label{fig:sketch}
\end{figure}

Our method for estimating basin stability is based on the observation that there are many large-amplitude perturbations in the Earth system, such as volcano eruptions, meteorite impacts and geoengineering, to name a few. When a perturbation displaces the Earth system out of its current state by a non-small amount, the system response contains information about the nonlinear stability of the perturbed system state, as illustrated in figure \ref{fig:sketch}.

For simplicity, perturbations are assumed to occur at random times, independently from each other, with random amplitudes. This is idealised as shot noise: waiting times between two perturbations are exponentially distributed, and perturbation amplitudes are drawn from a finite perturbation window, uniformly for simplicity. Given a time series of relevant system observables with total length $T_{tot}$, there are at least two relevant time scales in the dynamics: firstly the average waiting time between two perturbations $T_{pert}$, secondly the time that the system requires to converge to an attractor $T_c$. If these time scales are ordered as

\begin{eqnarray}
   \hspace{3cm} T_{tot} \gg T_{pert} \gg T_c,
\label{eq:ineq}
\end{eqnarray}
and if the perturbation amplitude is large enough for the trajectory to explore different basins of attraction in phase space, then a single (long) time series is sufficient for estimating the basin stability of different attractors. While technically challenging in a problem of realistic complexity, this can in principle be achieved by the following protocol:
\begin{enumerate}
    \item Estimate the location of attractors from the time series. An example of a method for finding fixed points from a low-dimensional time series is described in \cite{glover1993reconstructing}. Chaotic attractors may be reconstructed using the Takens Embedding theorem \cite{kantz2004nonlinear}.
    \item Identify disruptive events by finding isolated large jumps in the signal, akin to \cite{drijfhout2015catalogue}. Specifically, we search for points in the time series where the rate of change is much larger than at times immediately before and after. For the idealised shot noise scenario, given a sequence of observed values $\lbrace x_i\rbrace$ recorded at times $t_i =i\Delta t$, if the conditions
    \begin{equation}
        |x_{n-1}-x_{n-2}| \leq C |x_{n}-x_{n-1}| , \hspace{0.25cm} \land \hspace{0.25cm} |x_{n+2}-x_{n+1}| \leq C |x_{n+1}-x_{n}|,
    \end{equation}
    are satisfied with a small $0<C<1$, then we identify a jump at $t=t_n$. Numerically, we used $C=0.1$. For perturbations taking a finite time $\tau>\Delta t$ to occur, an adapted version of this procedure applies. We create list of jump times $t^*_j$.
    \item Subdivide the signal into segments between disruptive events, with each segment corresponding to the interval between two consecutive jumps, i.e. $[t^*_j,t^*_{j+1}]$.
    \item For each segment, check if the trajectory has converged to any attractor identified in step 1. For each attractor $A$, estimate $\mathrm{BS}(A)$ as the fraction of the number of trajectories $N_c(A)$ converged to $A$ divided by the number of all converged trajectories $N_c$:
    \begin{equation}
        \mathrm{BS}(A) \approx \frac{N_c(A)}{N_c}
    \end{equation}
\end{enumerate}
The system may further feature an underlying drift of a control parameter $\alpha$ with a third time scale $T_d$. If this drift is slow, such that $T_{tot} \gg T_d \gg T_{pert}$, then the above method may still be applied, if the time series is first split up into segments of length $O(T_d$), for each of which $\alpha$ is approximately constant.  

\section{Applications}
We illustrate the above procedure using two idealised models, where we compare our results against a ground truth established using the standard Monte Carlo simulation based estimation of basin stability \cite{menck2013basin}. 

\subsection{Amazon rainforest model}
\begin{figure}
    \centering
    \includegraphics[width=0.49\textwidth]{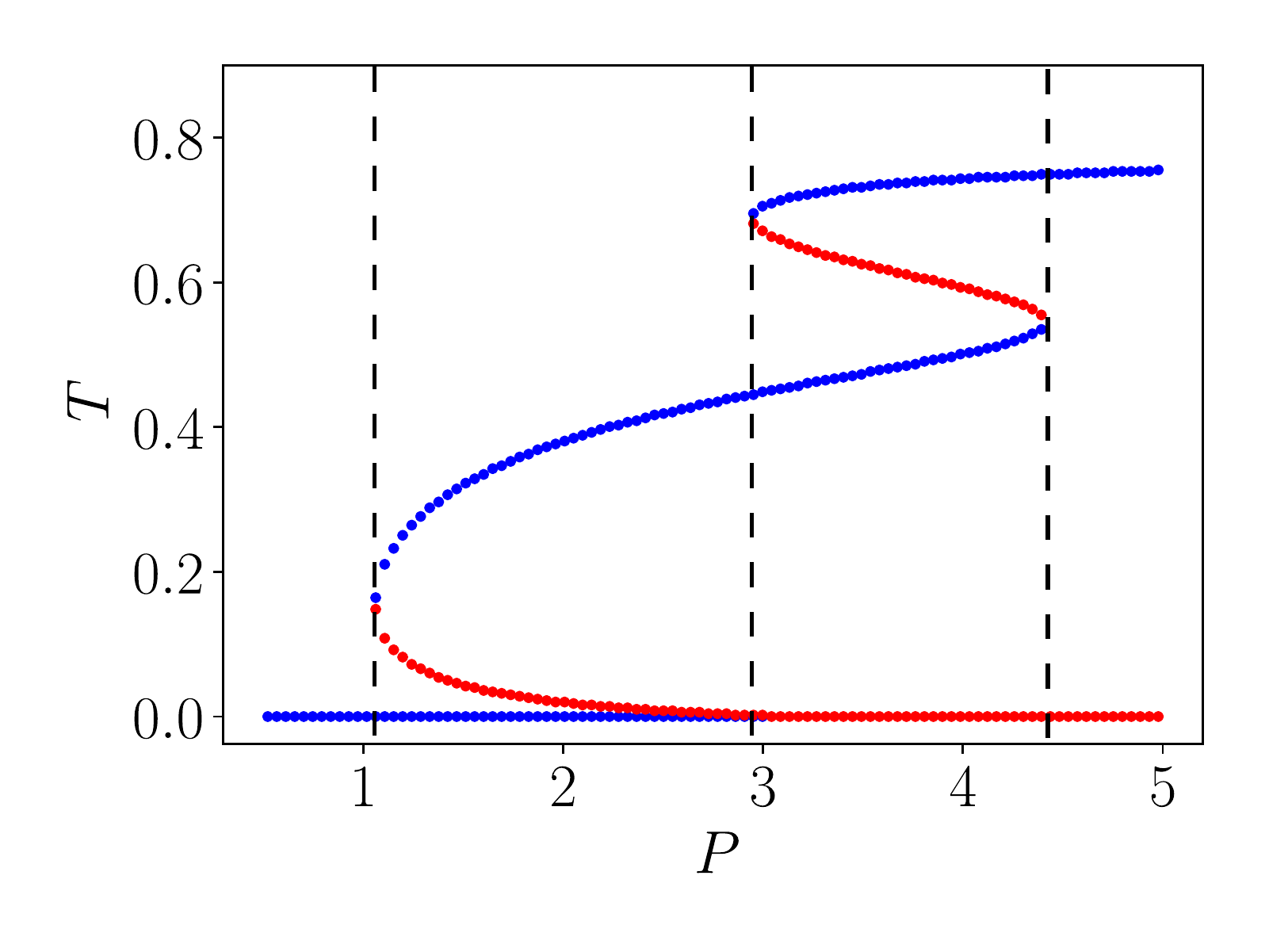}
    \includegraphics[width=0.49\textwidth]{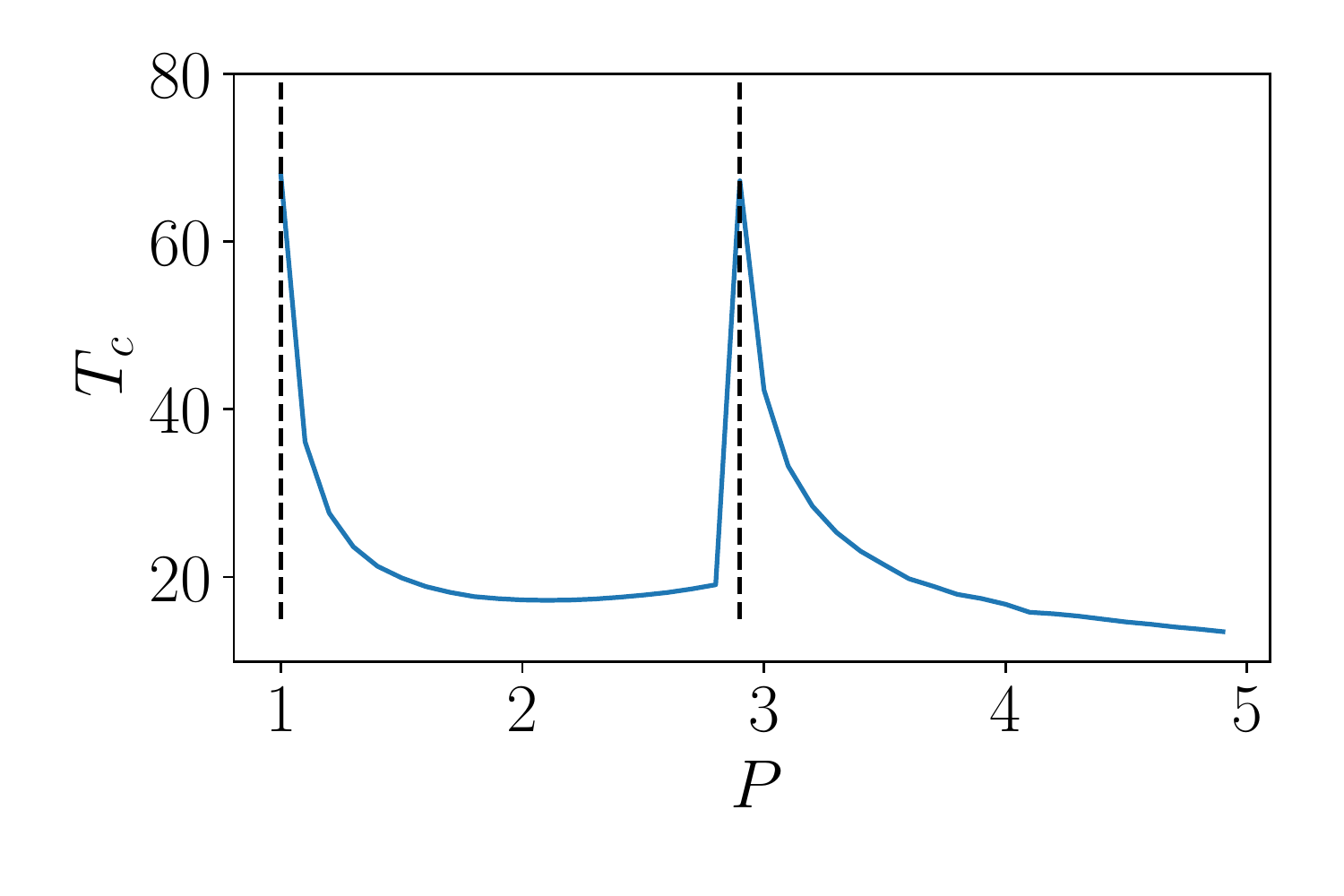}
    \caption{Left: Bifurcation diagram of the Amazon rainforest model by \cite{van2014tipping} with stable states in blue, unstable states in red. Right: convergence time scale toward attractor with highest tree cover versus daily precipitation $P$ (in $\mathrm{mm}/\mathrm{day}$), measured by exponential fit to approaching trajectories. A sharp increase is found near saddle-node bifurcations, signalling critical slowing down.}
    \label{fig:bif_diag}
\end{figure}

We first consider the simple one-dimensional Amazon rainforest model by van Nes et al. \cite{van2014tipping},
\begin{equation}
    \frac{dT}{dt} = r(P)T\left( 1- \frac{T}{k}\right) - m_A T \frac{h_A}{T+h_A} - m_f T \frac{h_f^p}{h_f^p+T^p}, \label{eq:vannes_ode}
\end{equation}
where $T$ is the relative tree cover $0\leq T \leq 1$, $t$ is time (in years), and the growth rate $r(P) = P r_m/(h_p+P)$. The parameter $P$ represents the mean annual precipitation in millimeter per day, which is an external control parameter in the model. All remaining quantities are constants, which are chosen exactly as in \cite{van2014tipping}, namely $h_A=0.1$, $h_f=0.64$, $h_p=0.5 \mathrm{mm}/\mathrm{ day}$, $K=0.9$, $m_A=0.15 \mathrm{yr}^{-1}$, $m_f=0.11 \mathrm{yr}^{-1}$, $p=7$, $r_m=0.3 \mathrm{yr}^{-1}$ (interpretations are given in Table 1 of \cite{van2014tipping}). As $P$ is increased from  $P =0.5 \mathrm{mm}/\mathrm{ day}$ to $P=5\mathrm{mm}/\mathrm{ day}$, the system undergoes a series of bifurcations, as shown in the left panel of figure \ref{fig:bif_diag}. Near those saddle-node bifurcations, the time required to converge to the stable equilibria increases rapidly due to critical slowing down. Figure \ref{fig:bif_diag} illustrates this for the states with the highest tree cover for a given $P$. The speed of convergence to an attractor naturally provides local (i.e. linear) information about the attractor's stability, a concept which has been termed \textit{engineering resilience} in the literature \cite{holling1996engineering}. \\

\begin{figure}
    \centering
    \includegraphics[width=0.49\textwidth]{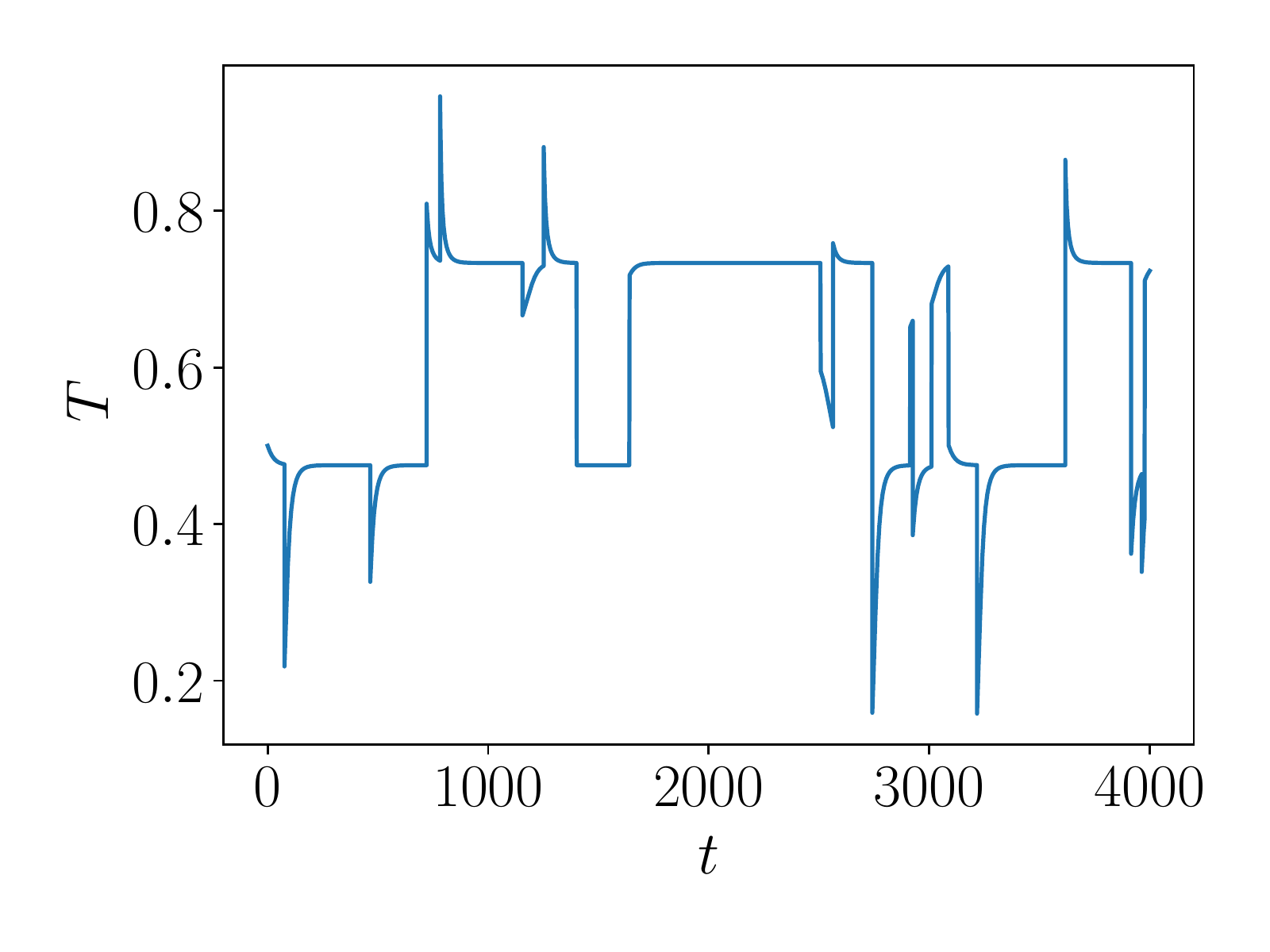}
    \caption{Sample time series of tree cover $T$ in the Amazon rainforest model generated from equation (\ref{eq:ineq}) including large perturbation for $T_{pert}=200 \hspace{0.05cm}\mathrm{yr}$ at $P=3.5 \mathrm{mm/day}$.}
    \label{fig:ts_vannes}
\end{figure}

Equation (\ref{eq:ineq}) is integrated in time from the initial condition $T(t=0)=0.5$ for fixed $P$ using a fourth-order Runge-Kutta scheme with time step $dt=0.1$. A random time $\tau$ is drawn from an exponential distribution $p(\tau) = \frac{1}{T_c} \exp(-\tau/T_c)$. After this time, a large-amplitude perturbation catapults the system into a new state, which is drawn, for simplicity, uniformly at random in $[0,1]$. As mentioned in section \ref{sec:method}, the source of the perturbations can be manifold, both human-induced or not. After the perturbation has taken place, a new $\tau$ is drawn and the procedure is continued until $t=T_{tot}$. A sample trajectory resulting from this procedure is shown in figure \ref{fig:ts_vannes}. \\

\begin{figure}
    \centering
    \includegraphics[width=0.48\textwidth]{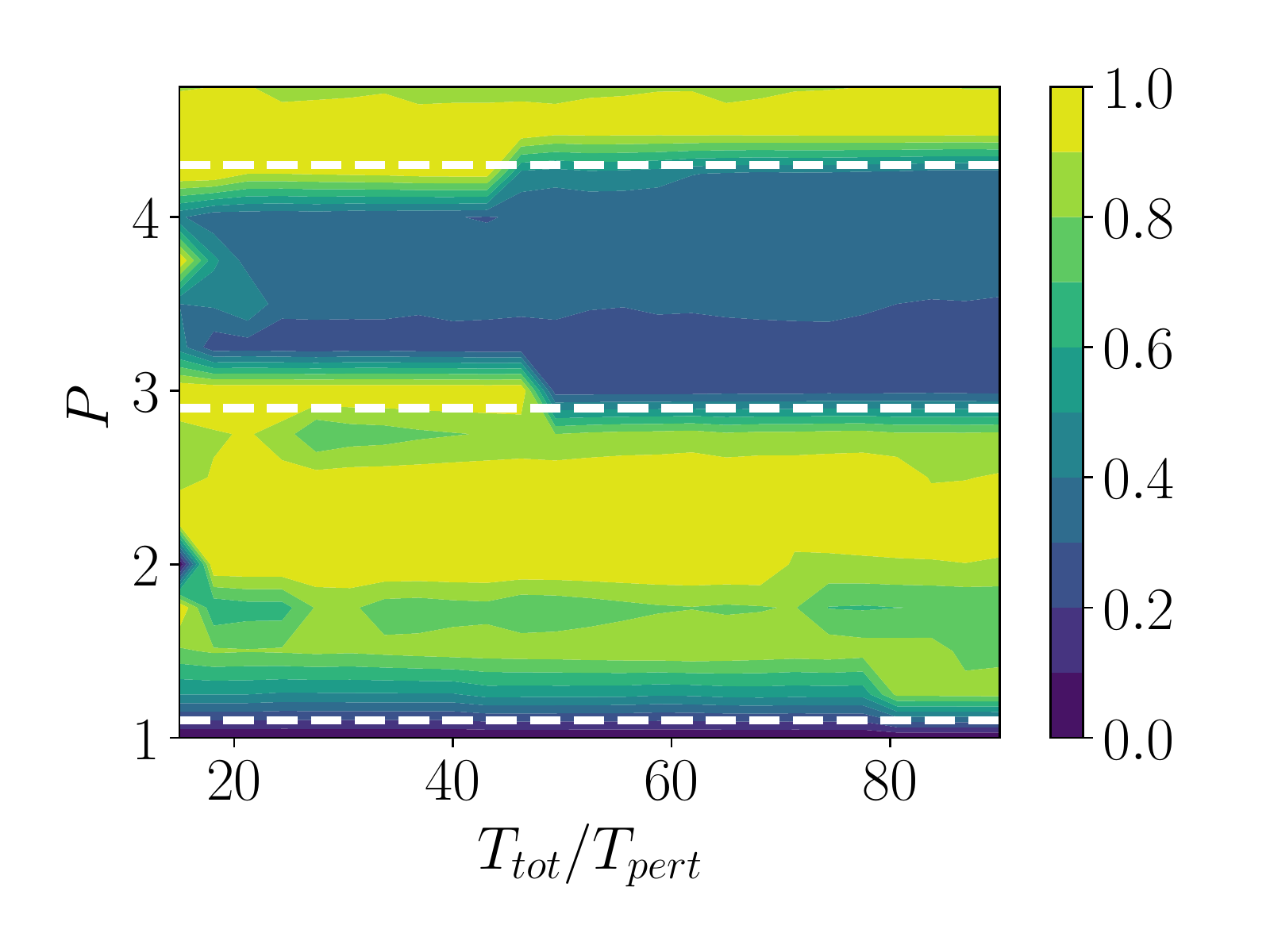}
    \includegraphics[width=0.48\textwidth]{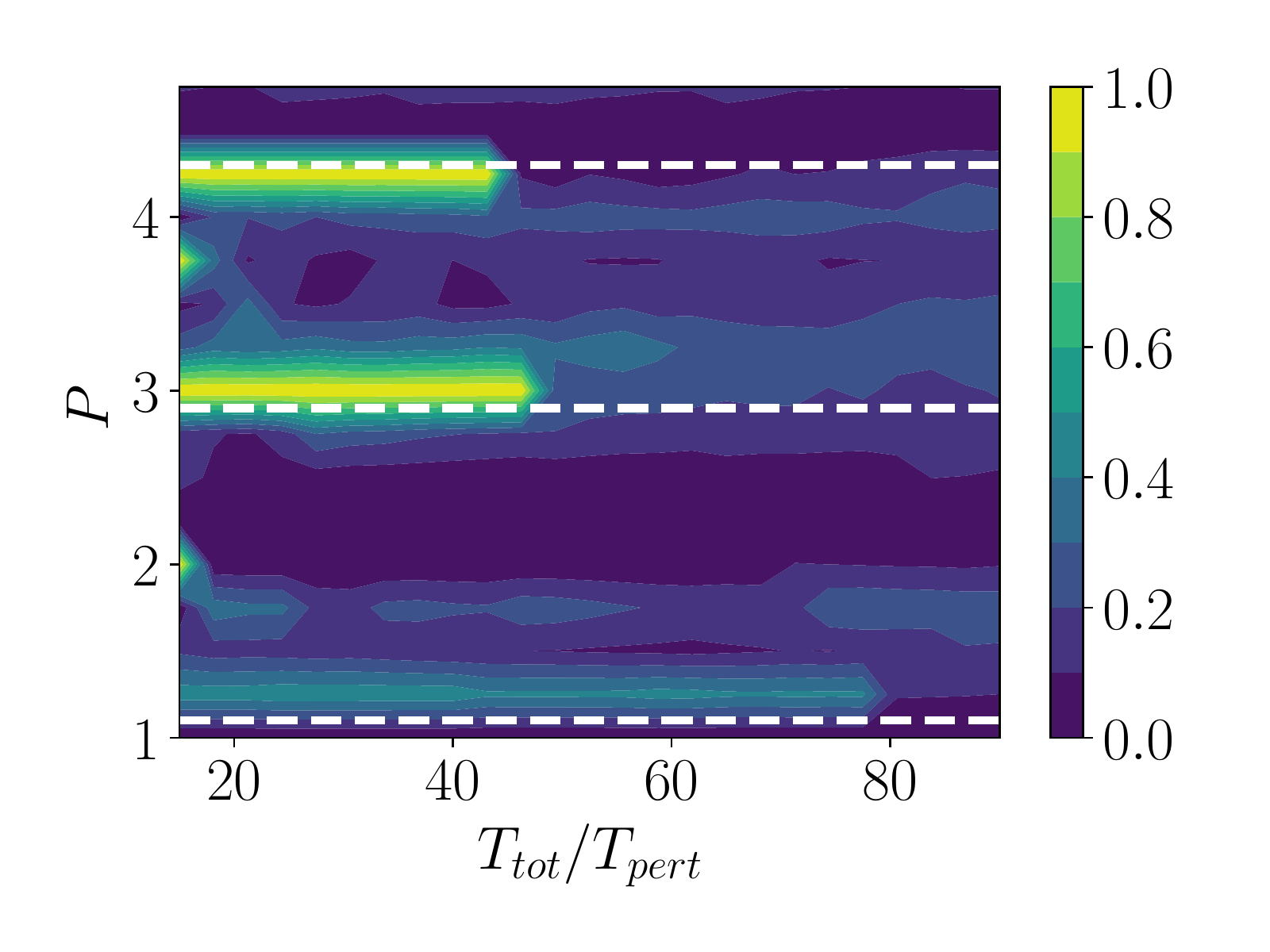}
    \caption{The quality of basin stability estimation improves with the number of observed perturbations and with distance from bifurcation thresholds.
    Left: basin stability of maximum tree cover state estimated from data {with varying number of observed perturbations $T_{tot}/T_{pert}$ and varying precipitation $P$}. Right: relative error as defined in eq. (\ref{eq:rel_err_def}). Relative errors larger than $100\%$ are shown as $1$. On the horizontal axis the total observation time is varied relative to the average waiting time between perturbations. The vertical axis shows the dependence on the precipitation parameter in $\mathrm{mm}/\mathrm{day}$. The waiting time between two perturbations is $T_{pert}=100  \hspace{0.05cm}\mathrm{yr}$ for all runs. According to figure \ref{fig:bif_diag}, this is much larger than $T_c$ away from bifurcations, but becomes comparable close to the threshold due to critical slowing down. Hence, it is expected there that convergence can be hindered by perturbations. Indeed, observed errors are largest close to the saddle-node bifurcations.}
    \label{fig:bs_vs_Tt}
\end{figure}

\begin{figure}
    \centering
    \includegraphics[width=0.48\textwidth]{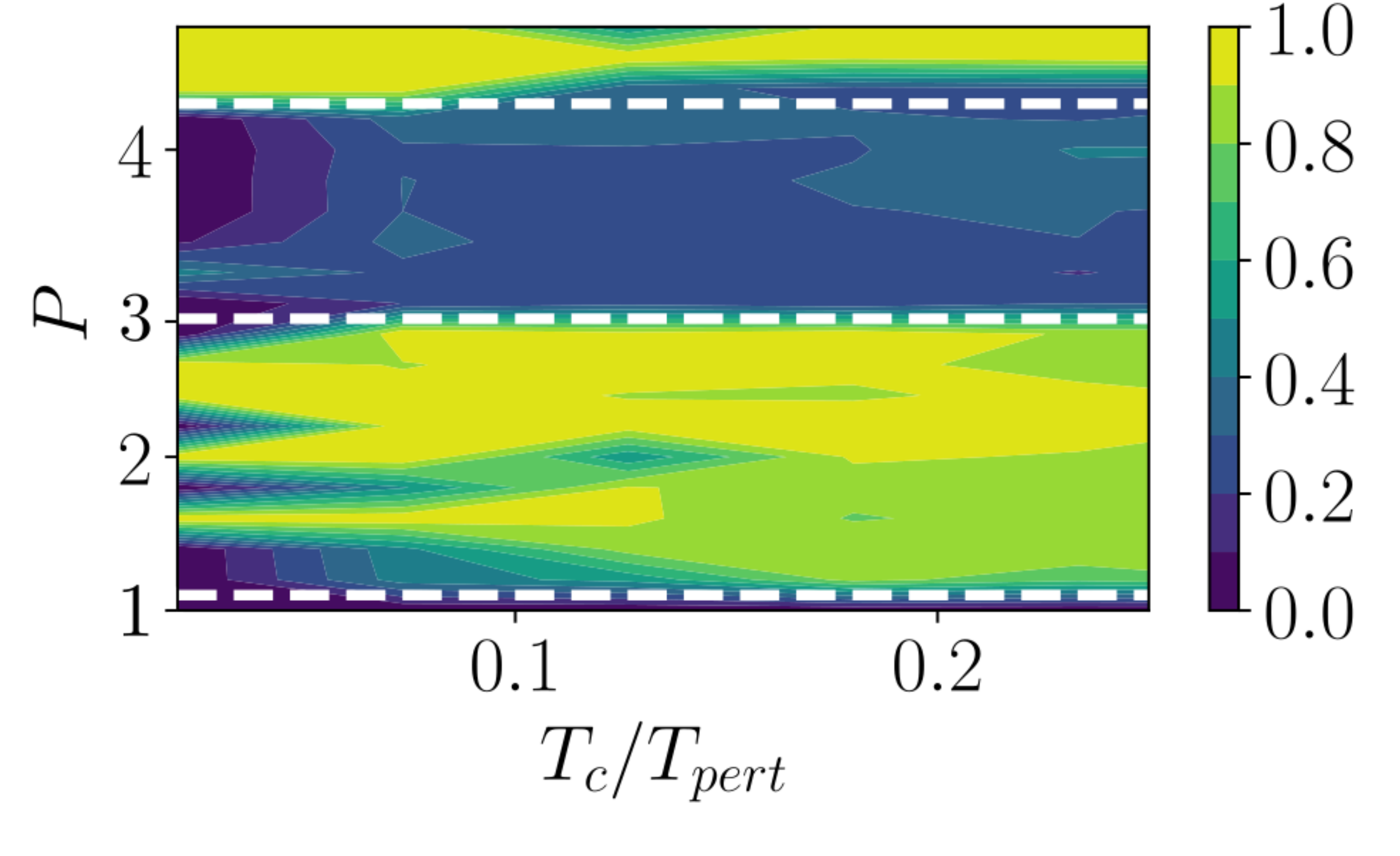}
    \includegraphics[width=0.48\textwidth]{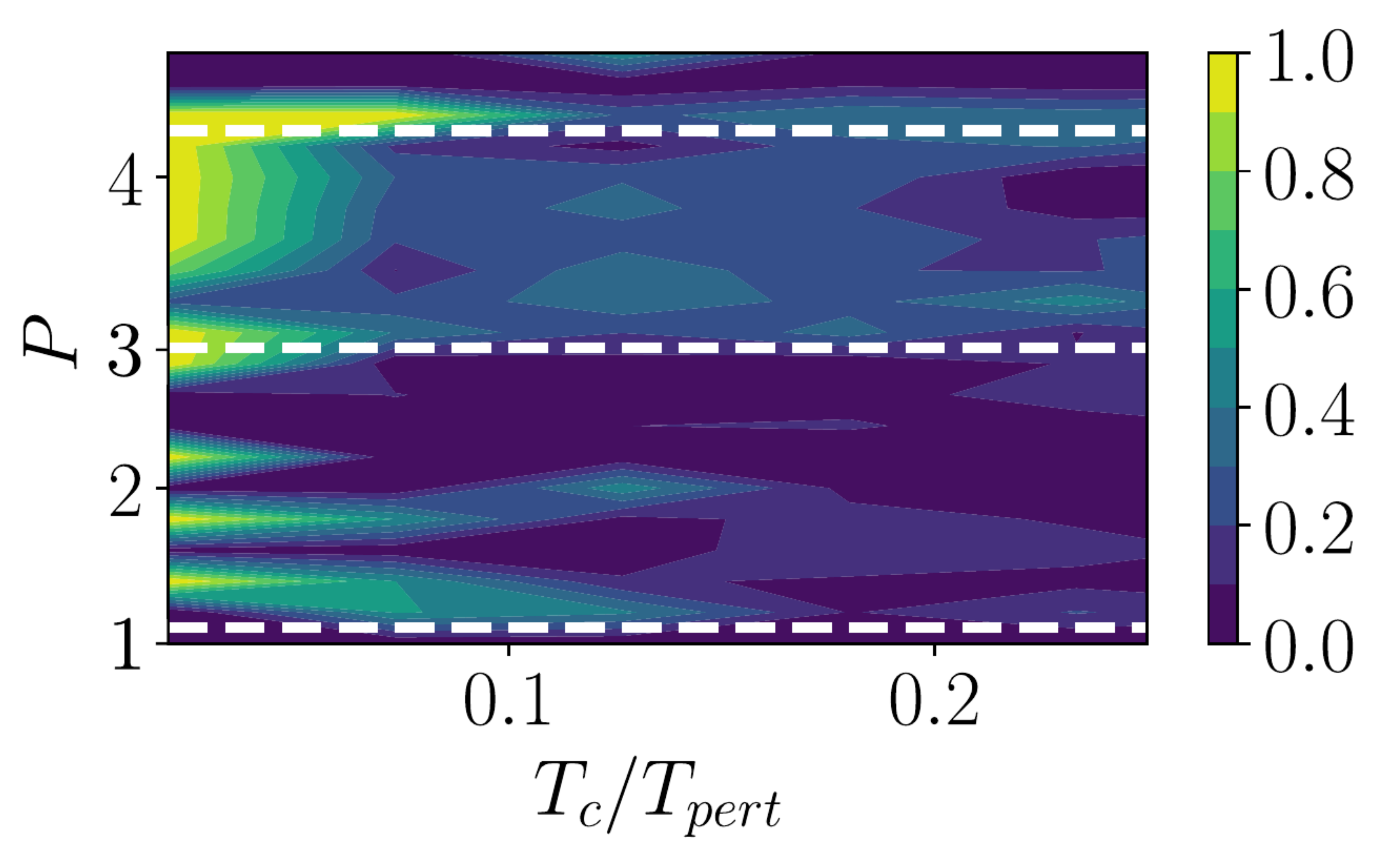}
    \caption{Left: basin stability of the maximum tree cover state estimated from data. Right: relative error as defined in eq. (\ref{eq:rel_err_def}). Relative errors larger than $100\%$ are shown as $1$. The horizontal axis shows $T_{c}/T_{pert}$, the ratio between convergence time mean return time of perturbations. The vertical axis shows the dependence on the precipitation parameter. The total observation time is $T_{tot} =40000\mathrm{yr}$ for all runs. }
    \label{fig:bs_vs_Tc}
\end{figure}

Once the trajectory has been generated, it is treated as an observed time series. The steps described in section \ref{sec:method} are applied to estimate the $\mathrm{BS}$ of the attracting state with the highest tree cover at a given $P$, by counting the number of perturbed trajectories converging to it. The fixed points are found from the signal based on sign changes of the derivative. Convergence is assessed by whether a trajectory has approached a fixed point to within $\pm\epsilon$ (we used $\epsilon=0.1$). The left panel of figure \ref{fig:bs_vs_Tt} shows the resulting estimated $\mathrm{BS}$ as a function of $P$ and $T_{tot}/T_{pert}$. The right panel shows the relative deviation 
\begin{equation}
 \hspace{3cm} \mathrm{err} = |\mathrm{BS}_{ex} - \mathrm{BS}_{est}|/\mathrm{BS}_{ex}, \label{eq:rel_err_def}
\end{equation} 
where the exact basin stability $\mathrm{BS}_{ex}$ is obtained by computing the exact fixed points of equation (\ref{eq:vannes_ode}) numerically, and $\mathrm{BS}_{est}$ is the estimate. For sufficiently long observation times, many perturbations have occurred and the errors in the estimation are small. The error is high, however, near the saddle-node bifurcations. This is simply a consequence of convergence to the equilibrium being slow there, such that the system cannot recover from a given perturbation before the next one occurs. This shows that it is hard to estimate nonlinear resilience close to a bifurcation threshold. Figure \ref{fig:bs_vs_Tc} shows the same quantities as figure \ref{fig:bs_vs_Tt}, but versus $T_c/T_{pert}$ and $P$, for fixed $T_{tot}$. Again, overall relative errors decrease with increasing {number of observed perturbations} ($T_c$ is fixed for a given $P$, so increasing $T_c/T_{pert}$ implies smaller $T_{pert}$, which implies larger $T_{tot}/T_{pert}$ at fixed $T_{tot}$), and the relative error is particularly pronounced near the bifurcation points.

\begin{figure}
    \centering
    \includegraphics[width=0.5\textwidth]{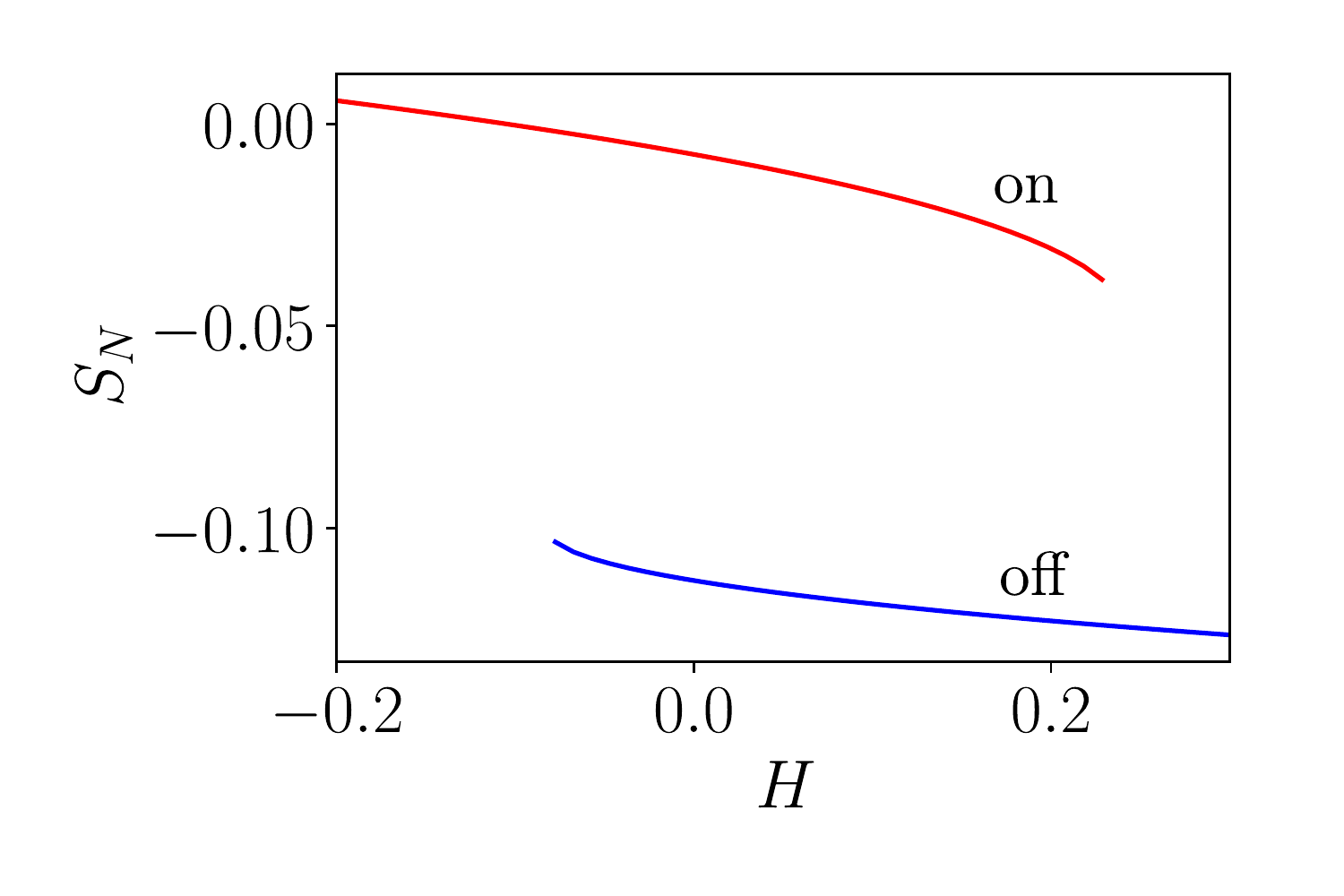}
    \caption{Stable branches constituting the hysteresis loop of the Wood et al. AMOC model as a function of freshwater hosing strength $H$. The AMOC's ``on'' branch is on top, and the ``off'' branch at the bottom.}
    \label{fig:bif_woods}
\end{figure}

\subsection{Thermohaline circulation model}

Now we consider the 5-dimensional ocean box model of Wood et al. \cite{alkhayuon2019basin}, which has been used successfully to investigate the multistability of the AMOC. For each of the five oceanic boxes in the model, labelled $i=N,T,S,IP,B$, there is a corresponding (rescaled) salinity perturbation $S_i$ relative to a base value $S_0$, that evolves in time due to fresh water forcing and coupling between boxes. A central quantity in the model is the AMOC flow $q$, given by 
\begin{equation}
    q= \lambda[\alpha(T_S-T_N)+0.01\beta(S_N-S_S)]
    , \label{eq:amoc_q}
\end{equation}
where $T_S,T_N$ are constant reference temperatures, and $\alpha,\beta,\lambda$ are model parameters. The latter, as well as the remaining parameters in the model are determined from the FAMOUS climate model and listed in appendix A of \cite{alkhayuon2019basin}. The salinity variables $S_i$ obey the nonlinear system of ordinary differential equations
\begin{equation}
    \frac{dS_i}{dt} = f_i\left(S_N,S_T,S_S,S_{IP},S_B,F_i\right), \label{eq:woodsetal}
\end{equation}
for $i=N,T,S,IP,B$, where $t$ is time in years and the nonlinear function $f_i$ is given in appendix A of \cite{alkhayuon2019basin}. It depends on the state variables as well as fresh water forcing $F_i$. The $F_i$ can vary due to varying hosing $H$, which we will consider our control parameter. It is defined such that the fresh water forcing in the Atlantic increases with increasing $H$, and decreases in the other ocean basins. For concreteness, we use 
$$    F_N= 0.383 + 0.1311 H, \hspace{1cm} F_T = -0.723 + 0.6961 H, 
$$
$$\hspace{0.1cm} F_S=  1.078 - 0.2626 H, \hspace{1cm} F_{IP} =-0.738 -0.5646H. $$
The $B$ box (bottom water) is not forced directly by fresh water in the model, i.e. $F_B=0$, and the remaining $F_i$ are such that\footnote{There is a minor typo in table 5 of \cite{alkhayuon2019basin},  since (\ref{eq:netforcing}) is not satisfied exactly with the values given there.}
\begin{equation}
    F_N+F_T+F_S+F_{IP}=0 \label{eq:netforcing}
\end{equation} Property (\ref{eq:netforcing}) ensures that the total salt content $ C=\sum V_i S_i, $ is conserved, with $V_i$ the volume of water in box $i$.

As $H$ is varied from $-0.2$ to $0.3$, the system undergoes several bifurcations, including two saddle-node bifurcations at $H\approx-0.1$ and $H\approx0.22$. There are two stable branches shown in figure \ref{fig:bif_woods}. A limit cycle is found in a very limited parameter range near $H\approx0.22$, but it does not lead to significant variations in the state variables $S_i$ and therefore does not affect our estimation procedure.  


As before, we generate a time series of length $T_{tot}$ by solving (\ref{eq:woodsetal}) using a fourth-order Runge-Kutta scheme, with perturbations occurring on average every $T_{pert}$ with exponentially distributed waiting times. We use the initial condition $\mathbf{S_0}=(S_N,S_T,S_S,S_{IP},S_B)=(-0,00879, 0.0435, -0,0573,-0.0573,-0,0462)$ corresponding to Table 3 of \cite{alkhayuon2019basin}. The perturbations $\Delta S_i$ are assumed to catapult the system to a new state $\mathbf{S}' = (S_N',S_T',S_S',S_{IP}',S_B')$, given by
$$ \mathbf{S}'= \mathbf{S_0} + (\Delta S_N,\Delta S_T,\Delta S_S,\Delta S_{IP},\Delta S_B)  
$$
with the $\Delta S_i$ drawn uniformly at random from $[-0.2,0.2]$ for $i=N,T,S,IP$, and $\Delta S_B$ ensuring that the total salt content remains unchanged. Although this perturbation window does not correspond to a physical scenario, it is suitable for establishing a ground truth basin stability for comparison with our estimates. Attractors are found based on low values of the derivatives. Jumps are identified from sudden increases in magnitude in the derivatives $dS_i/dt$, and basin stability is then estimated as the fraction of converged trajectories relaxing to a given attractor, following section \ref{sec:method}.

\begin{figure}[h]
    \centering
    \includegraphics[width=0.5\textwidth]{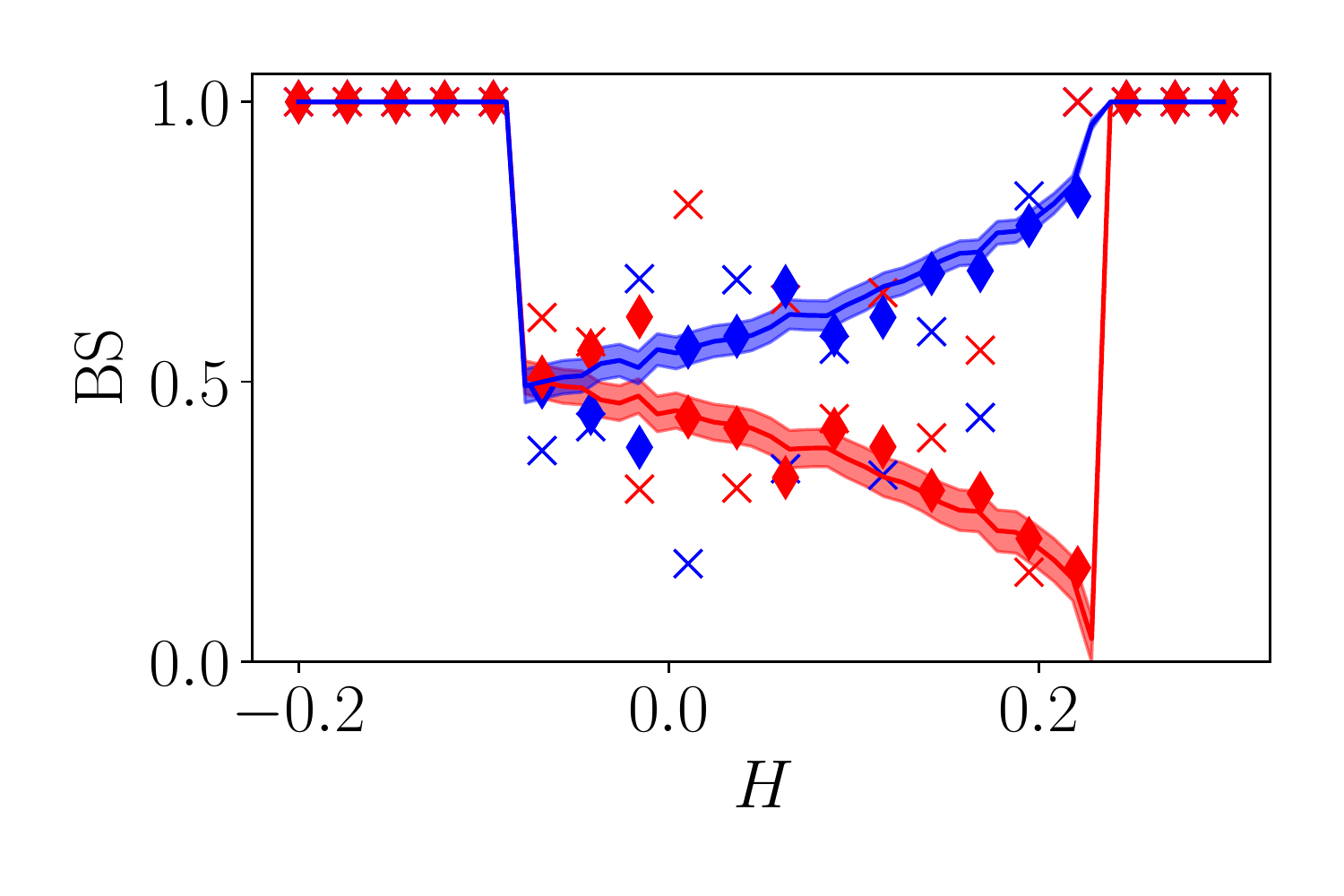}
    \caption{Basin stability in the Woods et al. model as a function of hosing $H$. Red: 'on' branch, blue: 'off' branch. Solid lines show ground truth basin stability relative to the perturbation window described in the text, with envelope corresponding to 3$\sigma$ from eqn. (\ref{eq:error_est}). Symbols show an estimation from a time series, with crosses corresponding to $T_{tot}/T_{pert} = 125$, and diamonds to $T_{tot}/T_{pert}=1000$. We used $T_{pert}=2000$.}
    \label{fig:bs_woods}
\end{figure}

Figure \ref{fig:bs_woods} illustrates the resulting estimate of basin stability for two different values of $T_{tot}/T_{pert}$, as a function of $H$, and compares it to the ground truth basin stability obtained from standard Monte-Carlo sampling in the same perturbation window. The error bar is obtained from the Bernoulli standard deviation
\begin{equation}
\sigma =  \sqrt{BS(1-BS)}/\sqrt{N}  \label{eq:error_est}
\end{equation}

\begin{figure}
    \centering
    \includegraphics[width=0.5\textwidth]{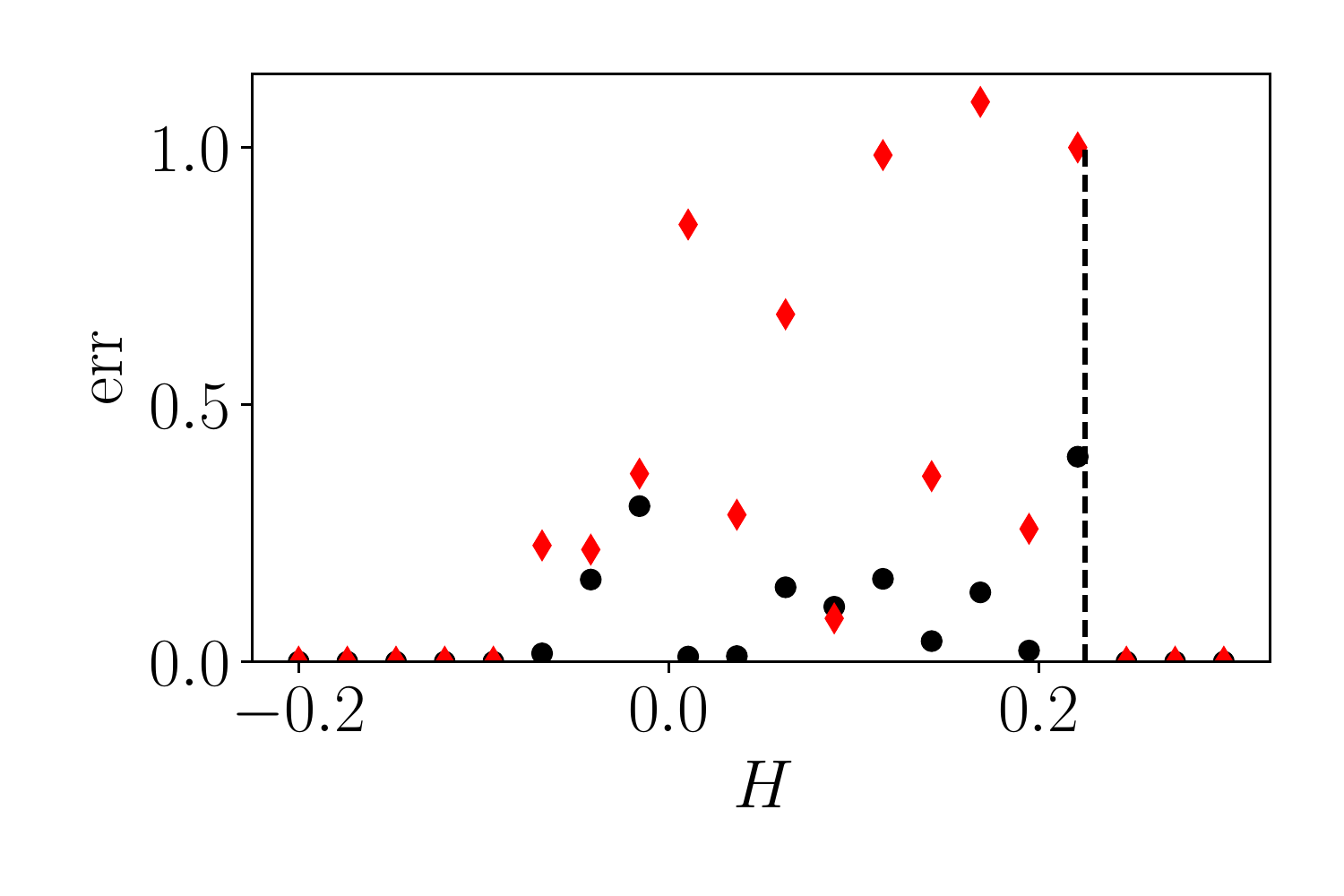}
    \caption{Large relative errors occur in the estimate of 'on' branch basin stability if observation time is too short. Vertical dashed line indicates end of on branch. Symbols show relative error of 'on' branch basin stability $\mathrm{BS}_{est}$ with respect to ground truth value $\mathrm{BS}_{ex}$, as defined in eq. (\ref{eq:rel_err_def}). Diamonds: for $T_{tot}/T_{pert}=125$ for each $H$, dots: $T_{tot}/T_{pert}=1000$ for each $H$. Longer observation times reduce relative error, but it remains significant in particular close to the end of the 'on' branch.}
    \label{fig:rel_err_ocean}
\end{figure}

At $_{tot}/T_{pert}=100$ for each value of $H$, there are still significant deviations in the estimate from the ground truth, as can be seen in figure \ref{fig:rel_err_ocean}. The figure also shows that the error can be reduced by increasing the observation time such that $T_{tot}/T_{pert}=O(1000)$, since more trajectories can be observed to converge to either attractor. Then, the estimate correctly captures the continuous decrease of $\mathrm{BS}$ with $H$, although the relative error is larger at the end of the 'on' branch, again due to slow dynamics and the small absolute value of the basin stability there.  Nonetheless, with a sufficiently long observation time, the estimate can serve as an early warning signal for the bifurcation.


\section{Conclusions}

In summary, we propose an alternative method for estimating global stability of a system, based on the analysis of long time series, and illustrate our approach in the context of two simple models. In the Amazon rainforest model of \cite{van2014tipping}, we showed that stability estimation becomes less reliable close to saddle-node bifurcation points due to critical slowing down. This can be circumvented by observing a large number of perturbations $O(100)$. Linear resilience measures can also capture the critical slowing down, and thus present a useful complement our estimation procedure close to bifurcation points. Away from bifurcation thresholds, the estimation was significantly more reliable even for lower numbers of observed perturbations. In the ocean box model by \cite{alkhayuon2019basin}, we illustrated that trends in basin stability can be captured, despite relative errors remaining significant near the end of the AMOC's on branch, and thus the estimate can potentially provide an early warning signal, provided that a large number of $O(10^3)$ perturbations had been observed for each hosing parameter value.

We have illustrated that the estimation of $\mathrm{BS}$ from time series data crucially depends on the ratios of at least three time scales in the problem: the total observation time, the mean waiting time between perturbations, and the intrinsic convergence time. Observations must be long enough for a sufficient amount of large-amplitude perturbations to have occurred, and the system must have time to relax between two perturbations, as expressed by condition (\ref{eq:ineq}). It would be of great interest to apply our approach to realistic observational (paleo)climate or ecosystem data, where parameters drift slowly \cite{steffen2018trajectories,westerhold2020astronomically}. In most cases, the observational data does not reach back long enough continuously to observe sufficiently many perturbations. In some cases, the observed perturbation amplitudes may not be large enough to reliably sample the phase space. A promising direction for further investigation are lake ecosystems \cite{su2021long}, where long observations are more readily available, and internal dynamics are fast compared to that of climate dynamics. For other systems, suitable data sets may not yet be available, but we believe that it would be very important to try to assemble such historical records allowing for an estimation of nonlinear resilience under realistic circumstances \cite{marwan2021nonlinear}. We leave this work for future studies.\\

A general difficulty one faces when attempting to apply our method to realistic time series data stems from the fact that one does not control perturbation amplitude in an observation. If perturbations are too weak to allow a comprehensive exploration of the state space, then the resulting stability estimates are biased, since the effective perturbation window does not cover the different basins of attraction. 
Realistic data, even when available over a sufficiently long time interval, obviously presents many challenges in addition to those explicitly discussed here. Despite the underlying dynamics typically being of high dimensionality, observations are usually only possible for a small number of system variables, such as the AMOC flow rate in the ocean. Perturbations could differ strongly between different system variables, and may thus be hard to detect. Furthermore, in addition to large-amplitude perturbations, observed signals are noisy and require non-trivial data treatment before yielding useful information about perturbations. This also makes it hard to reliably identify dynamical attractors.

To conclude, in view of the ever increasing amounts of observational data on the Earth system and its components in the era of big data \cite{Mahecha2020earth}, it will be of great benefit if nonlinear stability of attractors can be estimated directly from such observations. Despite the many complicating factors, we are convinced that this will be highly useful both to better understand what future changes and regime transitions lie ahead, and also to learn more about our planet's past. Our work presented in this paper is a first step in this direction.

\section*{Acknowledgements}
AvK acknowledges the support from Studienstiftung des deutschen Volkes and Ecole doctorale de Physique de l'Ile de France during his doctoral thesis work. AvK was supported in part by the National Science Foundation (grant DMS-2009563). AvK thanks Ann Kristin Klose for pointing out reference \cite{alkhayuon2019basin}. JFD was supported by the Leibniz Association (project DominoES) and the European Research Council (project ERA ``Earth Resilience in the Anthropocene"; grant ERC-2016-ADG-743080).

\section*{Author contributions}
All authors together conceived the study, AvK performed the simulations and drafted the paper, all authors reviewed the manuscript and contributed to the final version.

\section*{Conflict of interest}
The authors declare no conflicts of interests.

\section*{Data availability statement}
The data that support the findings of this study are available upon reasonable request from the authors.

\section*{References}
\bibliographystyle{unsrt}
\bibliography{biblio}

\end{document}